# Coupling phase-switching with generalized Brewster effect for tunable optical sensor designs

Daniel T. Yimam*, Dennis van der Veen, Teodor Zaharia, Maria Loi, and Bart J. Kooi*

Zernike Institute for Advanced Materials, University of Groningen, Nijenborgh 4, 9747 A Groningen, The Netherlands

d.t.yimam@rug.nl, b.j.kooi@rug.nl

## Abstract

The non-linear and tunable optical constants of phase-change materials associated with their phase-switching have been utilized in reconfigurable optical devices. For example, one possible application of phase-change thin films is for tunable perfect absorption designs, where p- polarized light reflectance vanishes at a specific incidence angle known as the Brewster angle. This work demonstrates a generalized Brewster effect (s- and p- polarized light absorption) for a multilayered heterostructure design based on the strong interference effect. The proposed design comprises a low-loss phase-change material, $Sb_2Se_3$, coated on a gold substrate. We experimentally and theoretically show the coexistence of vanishing reflectance values for both s- and p- polarized lights in the visible and near IR wavelength range at a single Brewster angle. Such vanishing reflectance values (points of darkness) are associated with phase singularities with abrupt changes. Moreover, we show that additional phase singularities can be realized by switching the active $Sb_2Se_3$ layer between amorphous and crystalline structures, extending the functionality of our design. The realized phase singularities are susceptible to small optical constant changes and can be utilized for ultra-sensing applications. As a proof of concept, we demonstrate our design's $CO_2$ gas sensing capabilities, showing a linearly dependent optical response with gas flow rate. We believe our lithography-free design with multiple phase singularities, from the coexisting generalized Brewster effect and from the additional structural switching, is highly promising for tunable optical sensing applications.

## Keywords

Phase singularity, Brewster effect, spectroscopic ellipsometry, gas sensor, optical sensor, phase change materials, perfect absorber

# Introduction

In nanophotonics and optoelectronics thin films play a key role in advanced optical device designs. By tuning the physical and optical properties of thin films, control of light interaction with optical devices has produced attractive applications in many fields. One important parameter tuned for such applications is the perfect absorption of the incidence light.[1,2] Applications in thin film solar cells and photovoltaics, photodetectors, sensors, and structural colors all benefited from smart optical device designs, achieving perfect light absorption within tunable spectral ranges.[3] One way of achieving perfect absorption is through light resonance with artificially constructed metal-dielectric-metal plasmonic nanostructures and antenna designs.[3–5] However, the heavy nanofabrication tasks needed to realize the specific nanostructured surfaces create practical limitations. Alternatively, lithography-free smart heterostructure designs can be considered for achieving perfect absorption through resonant enhancement in the Fabry Perot cavity[6–8] or strong interference formalisms.[1,9,10] One class of materials attractive for multilayered perfect absorption designs is phase-change materials (PCMs). Integration of PCM thin films in optical and optoelectrical devices has received great interest in the past few years. The fast and reversible phase-switching and the associated optical contrast between phases have huge promises for dynamically reconfigurable photonics and produced interesting results, including perfect absorption.[11,12] Phase-change-based perfect absorber designs offer flexibility in absorption tuning from the amorphous/crystalline reversible phase-switching.[13–17]

For optical systems with no transmitted light intensity (for bulk substrates, for example), perfect light absorption is translated to no reflectance. Such systems are explained by a point of darkness where no reflection intensity is detected for a specific polarized light. However, realizing a point of darkness requires selecting an appropriate wavelength range and light incidence angle. The point of darkness is highly useful since the associated non-trivial behavior of the light phase can be utilized for various applications.[18–21] In systems where the point of darkness is achieved, the phase singularities show a high sensitivity to a relatively small refractive index change and thus can be used for sensing purposes.[3,8,20,22,23] Although most reported results are from nanostructured plasmonic designs, the point of darkness can be realized as well through multilayered heterostructure stacks.[2] This is achieved by a well-known classical Brewster effect, where the Fresnel reflection coefficient for p-polarized light disappears at a specific incidence angle.[24] As a result of the vanished reflectance, phase singularities can form at the Brewster angle. Such systems have been exploited for sensing applications.[8,25] However, the Brewster effect falls short for various materials and designs where the assumption of nonmagnetic, isotropic, and homogeneous medium is altered. A more generalized Brewster effect is expected where vanishing reflectance will occur for both s- or p- polarized light.[26–29]

In this work, we experimentally and theoretically demonstrate the generalized Brewster effect for a multilayered heterostructure stack produced by pulsed laser deposition (PLD). We show that the strong interference effect can create perfect/strong light absorptions and associated phase singularities for both s- and p- polarized lights at a single Brewster angle. We constructed multilayered stacks with near zero reflection for both s- and p- polarizations realized at the same incidence angle (i.e., $\theta^s_B = \theta^p_B$). The coexistence of vanishing reflectance values produced separate phase singularities at the corresponding points of darkness. Our proposed heterostructure device consists of a thin lossy phase-change layer $Sb_2Se_3$ coated on a bulk gold substrate. $Sb_2Se_3$ is one of the new classes of low-loss phase-change materials that recently attracted research interest for multiple applications in programmable photonics and display devices.[30–33] We demonstrate here, for the first time to the best of our knowledge, that the phase singularities from the vanishing reflectance values coexist for a wide range of thickness values of the active $Sb_2Se_3$ layer. As a result, we fabricated an optical device with two phase resonances at a single light incidence angle of 70°. Our work demonstrates an innovative and lithography-free multilayered heterostructure design, where s- and p- polarized Brewster effects coexist at the same Brewster angle. Moreover, another set of phase singularities has been realized from the amorphous to crystalline phase switching of the $Sb_2Se_3$ layer, which is very promising for tunable sensing applications.

## Experimental

Multilayered heterostructures were produced by pulsed laser deposition (PLD). $Sb_2Se_3$ phase-change thin films with varying thicknesses were deposited on a gold substrate. A 10 nm spacer $LaAlO_3$ (LAO) layer was deposited between the gold and the active $Sb_2Se_3$ layer (mainly to prevent potential intermixing), and a 10 nm thick LAO capping layer was deposited on top. Electron beam evaporation (Temescal FC2000) was initially used to coat a 100 nm gold layer on a $Si/SiO_2$ substrate. The deposition parameters for $Sb_2Se_3$ were initially calibrated such that exact composition transfer (Sb40% - 60%Se) is achieved with precise thickness control. A 1 $Jcm^{-1}$ laser fluence, 1 Hz repetition rate, and 0.12 mbar of Ar processing gas were used for depositions.

All optical characterizations were performed by variable-angle spectroscopic ellipsometry (VASE, Woollam). Spectroscopic measurements were collected for individual layers where the optical constants of n and k of each layer in the multilayered heterostructure were extracted. For both as-deposited amorphous and crystalline phases of $Sb_2Se_3$, the $\psi$ and $\Delta$ parameters have been collected for multiple incidence angles. Data fitting was performed using the WVASE software, and the Tauc-Lorentz oscillator was used to model the optical constants of $Sb_2Se_3$. The extracted optical constants were used for reflectance and ellipsometry $\psi$ and $\Delta$ calculations. For the reflectance calculations, a transfer matrix (TM) formalism was used to model the multilayered stack. The Fresnel reflection coefficients for both s- and p- polarizations have been collected from individual interfaces.

To specifically choose an accurate thickness value of $Sb_2Se_3$ where the dual phase resonances exist, the reflectance calculations were first done for various thickness values. A heterostructure stack was fabricated after a single thickness value was selected. Angle-resolved and polarization-dependent reflectance measurements were performed using spectroscopic ellipsometry. The points of darkness and the phase singularities for both s- and p- polarized lights were realized theoretically and experimentally. The associated phase shifts and resonances were measured using ellipsometry parameters (the intensity ratio $\psi$ and phase difference $\Delta$). Moreover, we demonstrated our optical device's gas ($CO_2$) sensing capabilities by correlating the gas flow rate to the ellipsometry-measured parameters at the phase resonance points. A gas purging system, with a box enclosing the stage, was attached to the ellipsometry setup, as illustrated in Fig. 1c.

## Results and discussion

The proposed multilayered heterostructure device, the measured optical constants n and k of the active low-loss $Sb_2Se_3$ phase-change layer, and the schematic of the gas sensor experimental setup are presented in Fig. 1. The heterostructure device schematically shown in Fig. 1a is composed of a spacer LAO layer to prevent possible intermixing between the gold substrate and the $Sb_2Se_3$ phase-change layer, and capping LAO layer for protection, mainly against oxidation. Multiple thickness values have been considered for the $Sb_2Se_3$ layer but the LAO layers have a fixed thickness value of 10 nm. In Fig. 1b, the optical constants n and k of the active $Sb_2Se_3$ layer are shown for both the as-deposited amorphous and crystalline phases. An $Sb_2Se_3$ sample was pulsed laser deposited (PLD) on a $Si/SiO_2$ substrate to extract these optical constants. After collecting spectroscopic ellipsometry measurements in the spectral range of 350 – 1700 nm, the Tauc – Lorentz dispersion model was used to represent the optical constants of the $Sb_2Se_3$ layer. The measurements and data fitting was again done for the crystalline phase after annealing the sample for 5 min at 250 °C.

The optical constants of $Sb_2Se_3$ show interesting properties in the visible spectrum range. The index of refraction n has the highest contrast between the as-deposited amorphous and the crystalline phases in the visible frequency range, and the material's light absorption capabilities are limited to that range. To utilize the absorbing nature of the material in the visible range and the added degree of freedom from phase-switching, we focused our attention on the wavelength range of 350 – 1000 nm for gas sensing applications. The experimental setup for the gas sensing experiment is illustrated in Fig. 1c. A sealed box, with glass windows for light input from the source, and for reflected light collection by the detector covers the sample stage of the ellipsometer. The glass windows are placed such that measurement is only possible at a 70° incidence angle. The sealed glass has two tube openings: one for the gas inlet into the sample stage and another for exhaust. A pressure gauge controlled the gas flow rate.

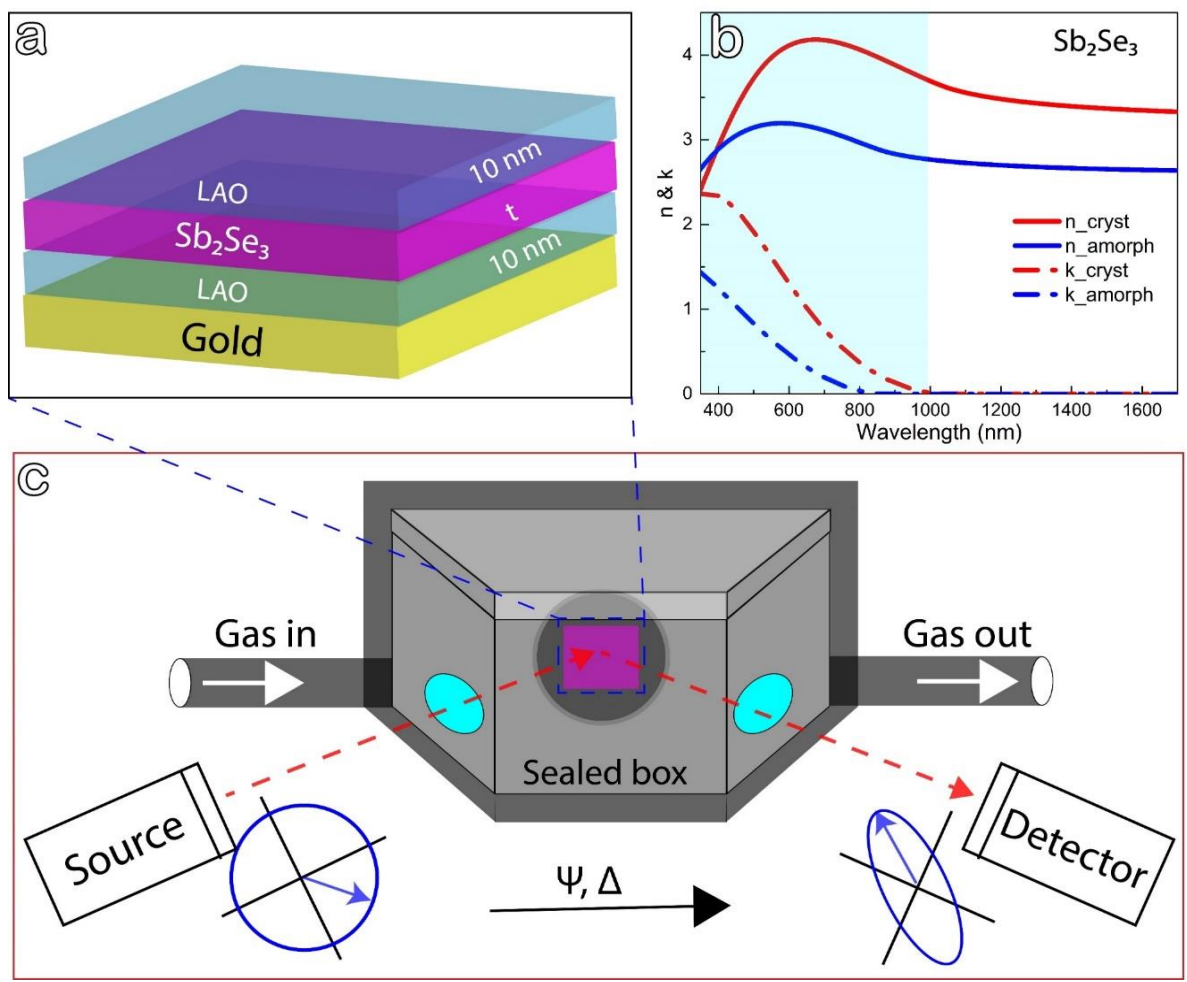

**Figure 1.** Schematic of a multilayered heterostructure device fabricated and the gas sensing experiment setup. (a) A phase-changing layer of $Sb_2Se_3$ was deposited on a gold substrate. Spacer and capping LAO layers were also deposited, each with a thickness of 10 nm. (b) The optical constants n and k of the active $Sb_2Se_3$ layer were extracted from spectroscopic ellipsometry measurements. (c) Schematic of the gas sensing measurement setup. The ellipsometry stage was covered by a sealed box with two glass window openings for input light from the source and output light directed toward the detector. The glass windows are located such that measurement is possible at 70° incidence angle. The sealed box has two openings, one for the gas inlet and one for the exhaust.

The reflectance values for both s- and p- polarized lights were calculated using the measured optical constants n and k of individual layers in the proposed heterostructure design. The optical constants were extracted from ellipsometry measurements and data fitting. The transfer matrix formalism was employed for the heterostructure stack presented in Fig. 1a, where the reflectance was acquired from the Fresnel reflection coefficients. Achieving a 'point of darkness' in multilayered heterostructure stacks is a relatively trivial task. The reason lies behind the large degree of freedom associated with parameter changes like layer thickness, wavelength range, and incidence angle for reaching the desired reflectance values. However, achieving points of darkness for both s- and p-polarizations is a difficult task. Therefore, our multilayered heterostructure design was carefully selected for various reasons. One of the main reasons, and the critical condition needed, is that perfect absorption of the polarizations (s- or p-) can be achieved in a certain spectral range with the proposed design. To theoretically investigate the reflectance responses, we calculated the reflectance of s- and p- polarized lights for the as-deposited amorphous $Sb_2Se_3$ layer thickness values of 0 – 45 nm (in steps of 1 nm) in the spectral range of 350 – 1000 nm. The angle of incidence was fixed at 70°. The calculated reflectance profiles and the ratios between s- and p- polarizations are presented in a color map in Fig. 2. The reflectance of p-polarized light in Fig. 2a and s-polarized light in Fig. 2b as a function of $Sb_2Se_3$ layer thickness show 'point-of-darkness' properties over large wavelength regions where vanishing reflectance values are seen.

For layer thickness values between 15-30 nm, there exist regions in the color maps (indicated by white arrow) where minimum reflectance values occur in the 350-500 nm region for the p-polarized light and in the 600-750 nm region for the s-polarized light. Interestingly, for the same $Sb_2Se_3$ layer thickness values in the range of 15-30 nm, we see a reflectance minimum for both s- and p- polarizations. The minimum reflectance values are associated with the Brewster effect. However, the observed Brewster effect differs from the classically known principle where only the p-polarized light vanishes. Our proposed design shows $Sb_2Se_3$ layer thickness and wavelength range combinations where both s- and p- polarized lights vanish. This effect is described by a new principle called the generalized Brewster effect. To pinpoint the location of minimum reflectance in terms of layer thickness and wavelength combinations, we calculated the ratios of s- and p- polarization reflectance values. The ratios are calculated as,

$$\Delta R_{s,p} = \left(\frac{R_{s,p} - R_{p,s}}{R_{p,s}}\right) \times 100\%$$

for all $Sb_2Se_3$ layer thickness values. The logarithm of ΔR values for both Rs/Rp and Rp/Rs cases are presented in Fig. 2c and 2d. As expected, extremely high ratio values are seen at the minimum reflectance values (maximum absorption).

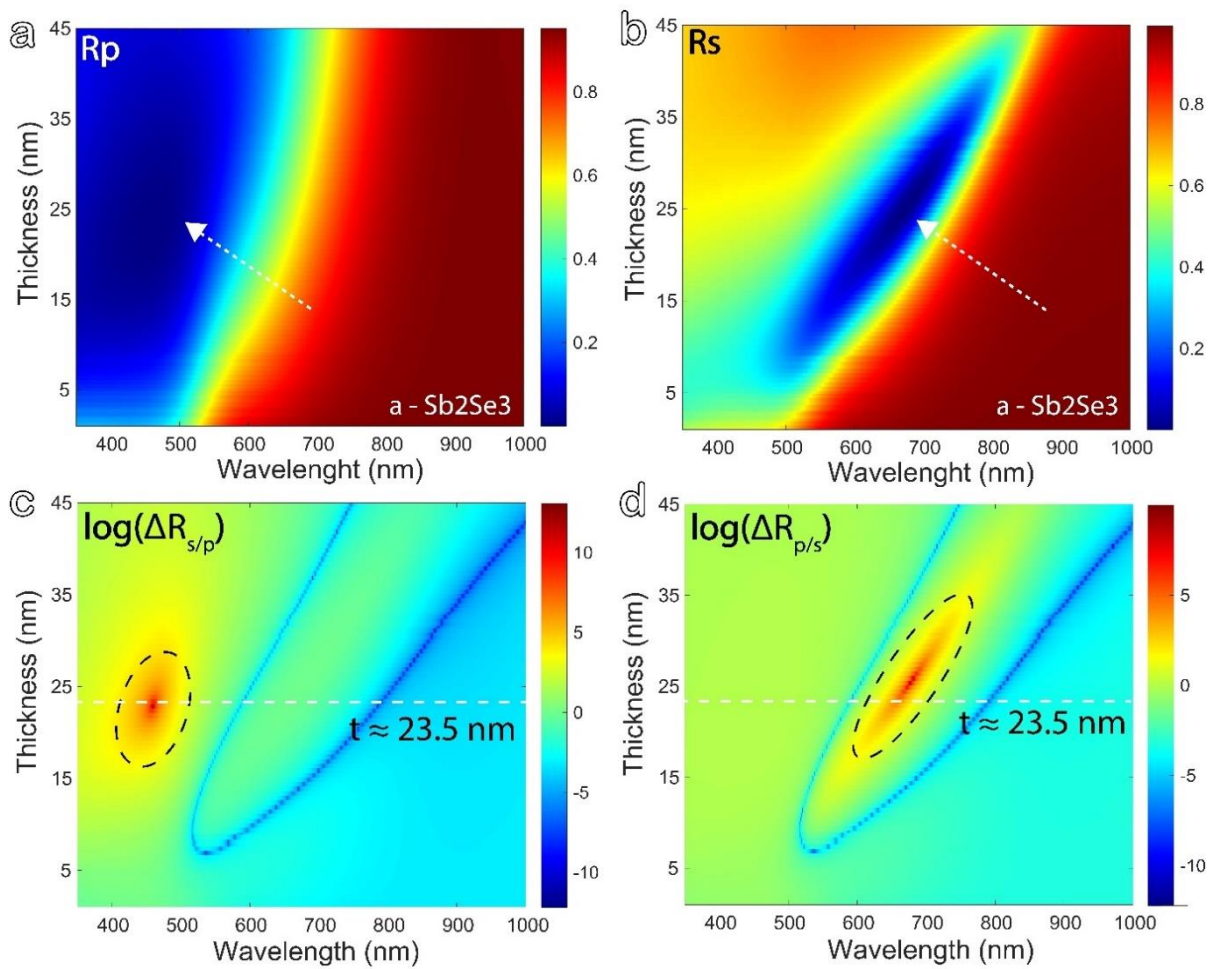

**Figure 2.** Calculated reflection vs. thickness results for the multilayered heterostructure device shown in Fig. 1a. The polarization-dependent reflectance profiles for (a) Rp and (b) Rs are computed for varying thicknesses of the active $Sb_2Se_3$ layer. The calculations are done for the spectral range of 350 – 1000 nm. Thickness values of low reflection (high absorption) are indicated for both s- and p-polarizations. The ratios of the polarizations (Rs/Rp and Rp/Rs) are calculated, and the log values are plotted in (c) and (d), respectively. The largest ratio values

were identified, and the associated thickness value was used to produce the reflective gas sensor device.

The ratio values in the color maps of Fig. 2c and 2d provide crucial information about our proposed multilayered heterostructure device and its possible application in gas sensing. In spectroscopic ellipsometry, the ψ and Δ parameters are directly related to the reflectance ratios of s- and p- polarized lights. Therefore, an abrupt phase change is expected at the maximum ratio values. Below is a detailed explanation of the ellipsometry parameters, reflectance ratio values, and singular phase behaviors. The maximum ratio values correspond to the minimum reflectance values observed in Fig. 2a and 2b. Moreover, for the layer thickness values of 15-30 nm, our calculated results show that the highest ratio values can be achieved at a single incidence angle value of 70° for both s- and p- polarized lights, and creates an opportunity to produce two resonance peaks in a single device. We fabricated a heterostructure stack based on the simulation insights to demonstrate the presence of dual resonances and phase singularities from the generalized Brewster effect. The multilayered stack was produced by PLD. For each layer, the pulse number to thickness conversion was carefully tuned such that sub-nm thickness control was possible (see also our earlier works with extreme thickness control for Sb and $Sb_2Te_3$ films).[34,35] To satisfy the existence of double resonances, from the s- and p- polarized Brewster effects at an incidence angle of 70°, we fabricated the stack with an $Sb_2Se_3$ layer thickness of ≈23.5 nm.

The reflectance values of s- and p- polarized lights were also calculated for the crystal phase of the active $Sb_2Se_3$ layer (see supplementary information (SI-1) Fig. S1a and S1b). Like the as-deposited amorphous phase, the crystalline phase also shows high absorption (vanishing reflectance) regions. Compared to the amorphous phase, the points of darkness for the crystalline $Sb_2Se_3$ phase are red-shifted (towards higher wavelength values). The minimum reflectance values are registered in the 500-600 nm region for the p-polarized light and in the 800-900 nm region for the s-polarized light. In addition, the reflectance ratio values are also calculated for the crystalline phase (see supplementary information (SI-1) Fig. S1c and S1d). Although the maximum ratio value for Rs/Rp and Rp/Rs shifted towards smaller and larger $Sb_2Se_3$ layer thickness values, a significantly large ratio value can still be achieved at a layer thickness of 23.5 nm. This presents a crucial opportunity where the points of darkness can be tuned through the $Sb_2Se_3$ amorphous-crystalline phase-switching without compromising the phase singularities. More details on this are given below.

For the fabricated multilayered stack, the measured and calculated reflectance spectra of the s- and p- polarized lights and the relative ratios between them are presented in Fig. 3. Keep in mind that, for our multilayer stack, the Brewster angle for both polarizations is 70°. Thus the reflectance measurements and calculations are done at 70°. Reflectance values measured with variable angle spectroscopic ellipsometry and calculated using the transfer matrix formalism are presented in Fig. 3a and 3b, respectively. The linear and log reflectance values are given for both the s- and p- polarizations. The reflectance curves show that the generalized Brewster effect is achieved where a reflectance minimum for both polarizations is demonstrated. In both Fig. 3a and 3b, the log reflectance curves (dotted red and blue curves) pinpoint the points of darkness where the s- and p- polarized lights vanish. Despite the small shift in reflectance and wavelength values, the measured reflectance results in Fig. 3a match the calculated values in Fig. 3b very well. The small discrepancies between the measured and calculated values are probably caused by the assumption of 'perfectly matching interfaces' in the calculations whereas fabricated samples have not such ideal interfaces. Nonetheless, small reflectance values of $R_p$ = 0.3% and $R_s$ = 3.8% are measured experimentally at the two points of darkness, more than enough to produce phase singularities.

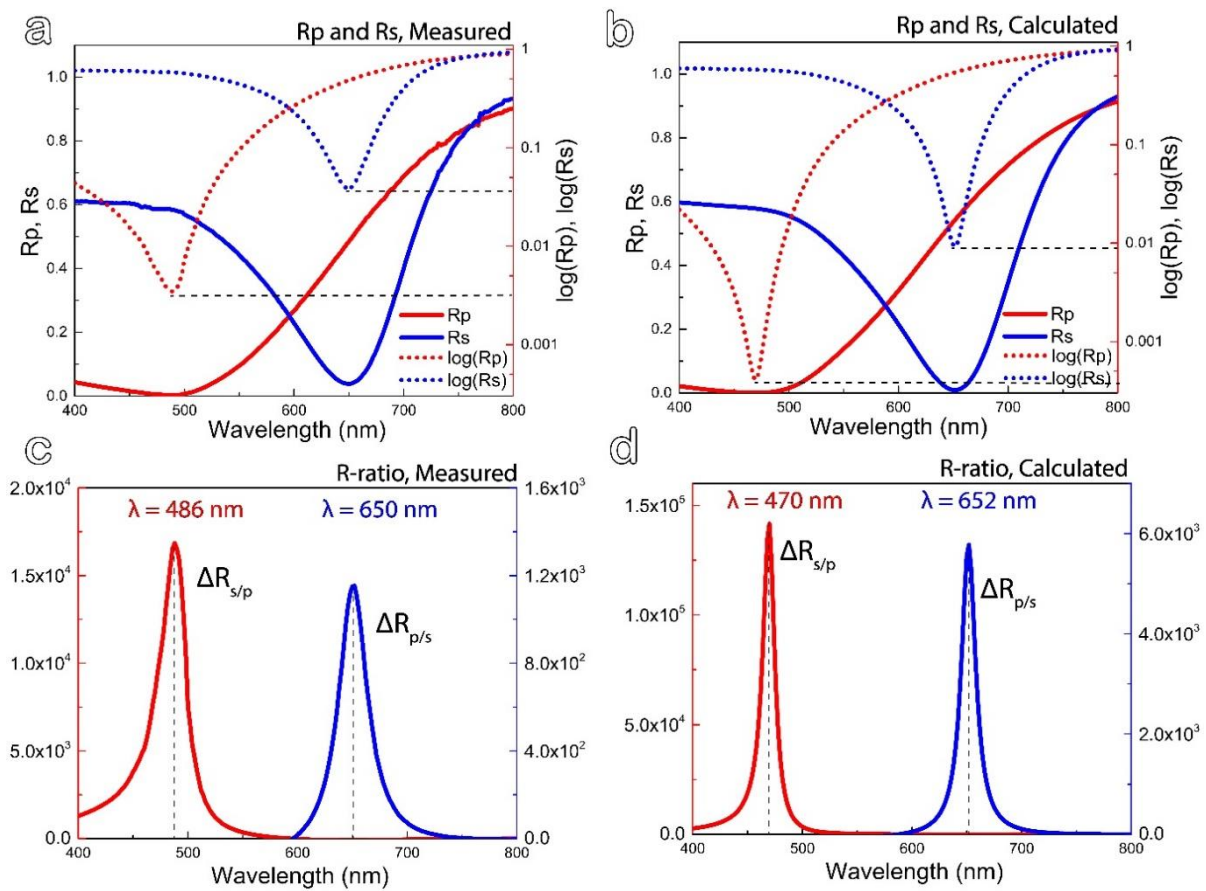

**Figure 3.** Measured and calculated polarization-dependent reflectance profiles for a multi-layered heterostructure of 10 nm LAO/23.5 nm $Sb_2Se_3$/10 nm LAO/100 nm gold/$SiO_2$. The (a) measured and (b) calculated reflectances for both s- and p- polarizations, at 70º incidence angle, are presented. In both (a) and (b), the right column presents the log reflectance values, indicating the point of darkness/high absorptions. The relative reflectivity ratios between polarizations for both measured and calculated values are presented in (c) and (d).

The ratios of the reflectance values from the s- and p- polarized lights, for both measured and calculated values, are presented in Fig. 3c and 3d. The two resonance peaks correspond to the s- and p- polarizations point-of-darkness. Comparable measured and calculated wavelength values of ≈ 650 nm can be seen for ΔRp/s. However, the measured values for ΔRs/p are red-shifted to 486 nm compared to the calculated wavelength value of 470 nm. As mentioned above, the two resonance peaks corresponding to the reflectance minimum points or points of darkness for both s- and p- polarized lights directly correlate to distinct phase behaviors where abrupt changes are expected. We measured and calculated the ellipsometry parameters ψ and Δ for the spectral range of 350 – 800 nm at 70° angle of incidence to confirm the presence of the phase singularities. The measured and calculated ψ and Δ parameters are presented in Fig. 4. For better correlation of the effect of points of darkness in the ellipsometry parameters, Rs and Rp values are also included in the figures.

To understand the effect of amorphous to crystalline phase-switching, we calculated the s- and p- polarization reflectance values for our heterostructure stack with a 22 nm and 23.5 nm $Sb_2Se_3$ layer in the as-deposited amorphous and crystalline phases (see supplementary information (SI-2) Fig. S2a). The 22 nm $Sb_2Se_3$ thickness value accounts for the thickness reduction upon crystallization. Phase-change materials show an increase in density when switching from amorphous to crystalline phases (usually between 6 – 10 %). This density change will be accounted for by a thickness reduction. More information for the 22nm thick layer is provided in supplementary information (SI-1 and SI-3). Compared to the amorphous film, the points of darkness for both polarizations are red-shifted towards higher wavelength values for the crystalline film. In Fig. S2b, the reflectance ratio values clearly show two resonance peaks (at 584 nm and 820 nm for Rs/Rp and Rp/Rs, respectively) corresponding to the vanishing reflectance values in s- and p- polarized lights. Since the reflectance ratio resonance peaks are associated with phase singularities, we calculated the ellipsometry parameters ψ and Δ for the crystalline $Sb_2Se_3$ phase. Indeed, the existence of these singularities are confirmed also for the crystalline $Sb_2Se_3$ layer in Fig. S2c and S2d from the calculated ψ and Δ parameters.

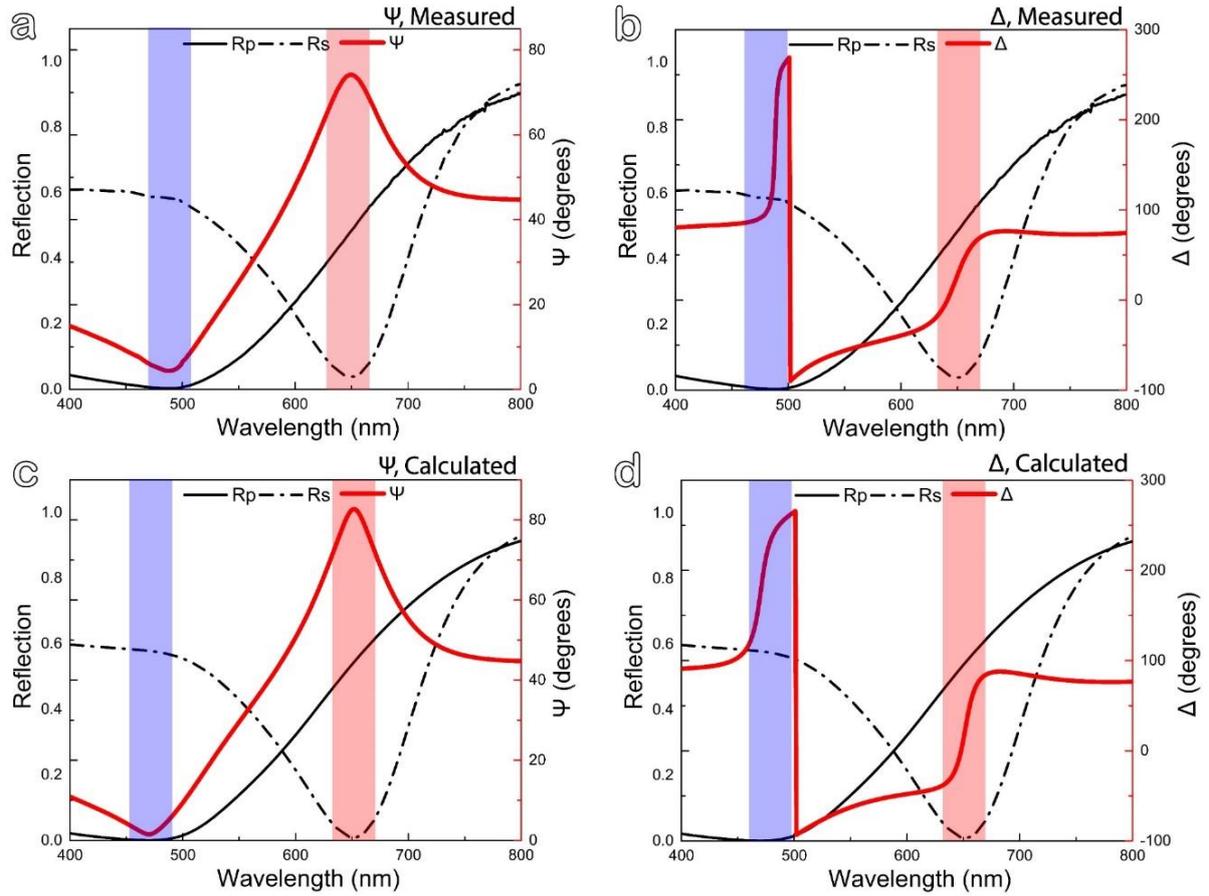

**Figure 4.** Measured and calculated ellipsometry parameters ψ and Δ overlayed with s- and p- polarization reflectance values for a multilayered heterostructure of 10 nm LAO/23.5 nm $Sb_2Se_3$/10 nm LAO/100 nm gold/$SiO_2$. The measured (a,b) and calculated (c,d) ψ and Δ values show minimum/maximum values and phase singularities at both s- and p- polarization point of darkness locations.

The measured (in Fig. 4a and 4b) and calculated (in Fig. 4c and 4d) ellipsometry parameters ψ and Δ match very well. However, at the specific wavelength values where the points of darkness were achieved, some distinct properties can be seen for both ψ and Δ parameters. The singular characteristics of ψ and Δ are directly related to the reflectance ratios of s- and p- polarized lights, thus the resonances seen in Fig. 3. In ellipsometry, when the initially linearly polarized light is reflected off of a sample at a non-zero incidence angle, the intensity and phases of the s- and p- polarized lights will change. For each polarization, the light propagation through the materials and the collected intensities from each interface is explained by the Fresnel equations. More importantly, the reflection ratio of Rp/Rs, represented by the complex reflection coefficient ρ, correlates directly to the ellipsometry ψ and Δ parameters through the equation $\rho = \frac{Rp}{Rs} = \tan(\psi) \exp(i\,\Delta)$. The equation clearly shows how ψ and Δ could be directly affected by the Rp/Rs ratio change. It explains the singular characteristics of ψ and Δ exactly at the minimum Rs and Rp reflection values at the points of darkness.

From Fig. 4a and 4c, we notice that the ψ parameter approaches 0° when Rp →0 and 90° when Rs →0. The singular modulation of ψ at the points of darkness is due to the relationship of ψ with the complex Fresnel coefficients $r_p = |r_p|\exp(i\varphi_p)$ and $r_s = |r_s|\exp(i\varphi_s)$ through the equation $\tan(\psi) = |r_p|/|r_s|$. For our multilayered heterostructure stack, ψ = 4.39° at 486 nm and ψ = 74.17° at 650 nm, consistent with the theoretical basis. Moreover, extreme singularities at the points of darkness are observed from the phase difference between the two polarizations (i.e., $\Delta = \varphi_p - \varphi_s$). The abrupt jumps in the Δ parameter are also clearly visible in Fig. 4b and 4d. Considering the measured values, the Δ parameter changes from ≈90° to ≈250° and from ≈-20° to ≈70° near the two points of darkness associated with the reflectance minimums. Note that the range of the abrupt change in Δ is larger (≈160°) when Rp →0 compared to when Rs →0 (≈90°). This is due to the relatively higher Rs value of 3.8% and provides an important insight into designs where larger Δ changes are achieved in the case of perfect absorption. The abrupt ψ and Δ parameter changes at the points of darkness are susceptible to extremely small optical deviations. Therefore, these phase shifts could be utilized for various sensing applications.

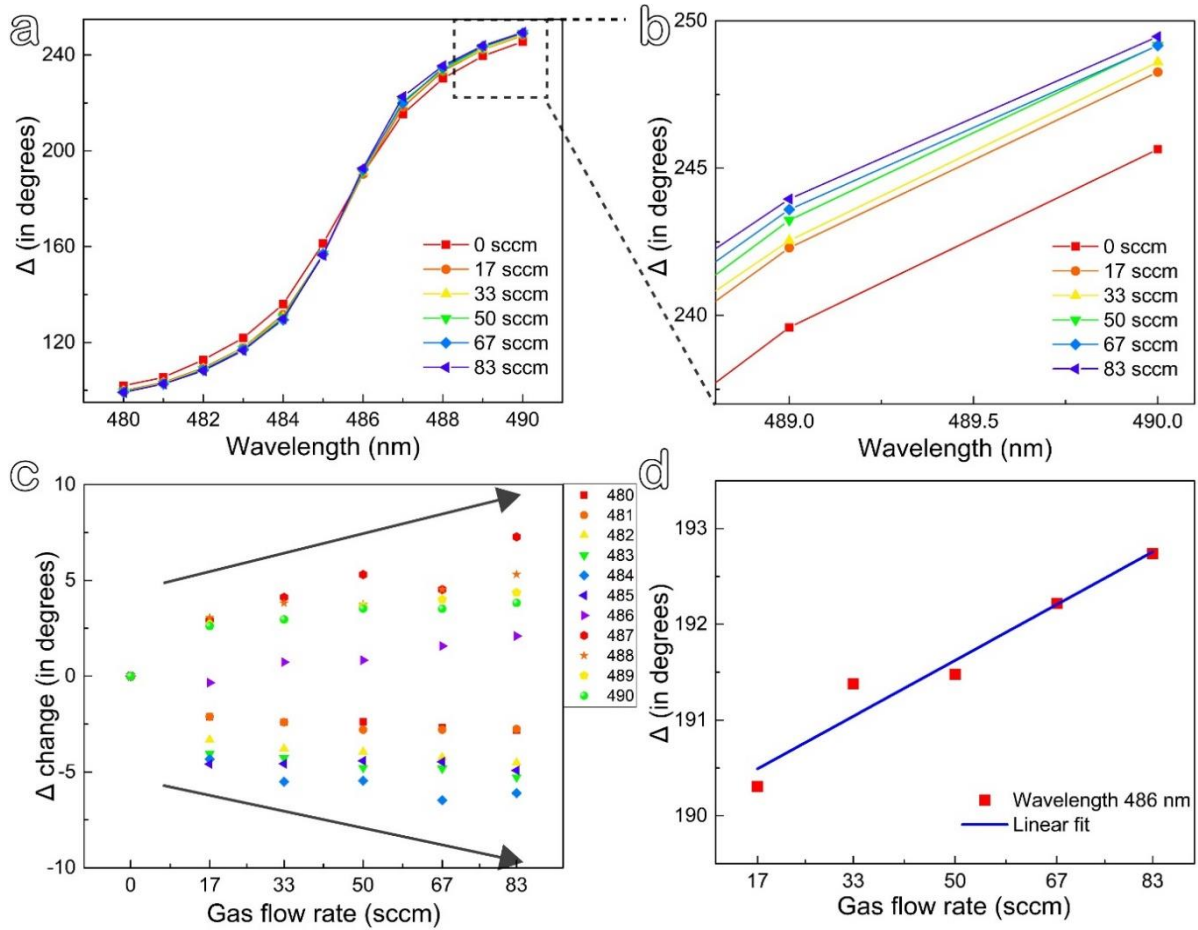

**Figure 5.** Results from gas sensing experiment. The flow rate of $CO_2$ gas was changed linearly, and ellipsometry $\psi$ and $\Delta$ parameters were collected. (a) The $\Delta$ value was collected for various flow rates in the 480 – 490 nm spectrum range. (b) A zoomed-in image of some of the $\Delta$ values presented in (a) shows a systematic trend between $\Delta$ and the $CO_2$ flow rate. (c) The relative difference in $\Delta$ values (relative to $\Delta$ value with no gas flow) shows a linear dependence on the gas flow rate. (d) An example of a linear correlation of $\Delta$ value change with gas flow rate at $\lambda$ = 486 nm, i.e. the measured point of darkness.

As a proof of concept, we employed our multilayered heterostructure stack for sensing extremely small $CO_2$ gas concentrations. As illustrated in Fig. 1c, a $CO_2$ gas canister was attached to the ellipsometry setup. The ellipsometry parameters were measured at the first resonance point (between 480-490 nm) while linearly changing the gas flow rate (from 0-83 sccm). Once the gas flow started, we waited 5 min so that the flow stabilized. Then, high-accuracy measurements were done for all flow rates where the polarizer/compensator combination was used to collect measured data. Moreover, five measurements were done in sequence for each flow rate, and the average value was taken as the final result. The results from the gas sensing experiments are presented in Fig. 5 only focusing on the $\Delta$ values. The completely analogous figure of Figure 5 but now for $\psi$ instead of $\Delta$ is shown in supplementary information (SI-4) Fig. S4. Fig. 5a shows the $\Delta$ values for all gas flow rates and a zoomed-in view on a small wavelength region is presented in Fig. 5b. The measurements show a shift toward higher Δ values for wavelengths higher than 285 nm and lower Δ values for wavelengths lower than 485 nm. To better visualize the Δ value changes to gas flow rate, we plotted the relative Δ value changes in Fig. 5c. The Δ change values are calculated by subtracting the initial zero gas flow measurement from each measurement. A clear upward trend for wavelength values above 485 nm and a downward trend for the rest is visible. The Δ changes show a somewhat linear dependence on the gas flow rate changes. Although relatively small, such linear correlation is also seen in the measured ψ value changes (see supplementary information (SI-4) Fig. S4). In Fig. 5d, an example of this linear Δ value change for a specific wavelength value of 486 nm is presented as an example.

Our work demonstrates a generalized Brewster effect where perfect/strong light absorptions were realized for both the s- and p- polarizations for a single Brewster angle. A lossy phase-change layer $Sb_2Se_3$ layer was coated on a gold substrate. Vanishing reflectance values for both s- and p- polarizations were achieved at an incidence angle of 70° by fine-tuning the active $Sb_2Se_3$ layer thickness for the wavelength range of 350 – 1000 nm. Our realized multilayered stack, capable of double darkness points associated with s- and p- polarized lights, can produce rapid phase shifts, as demonstrated experimentally and theoretically. Moreover, additional sets of phase singularities have been realized from the amorphous to crystalline phase-switching of the active $Sb_2Se_3$ layers. Such robust and tunable device designs, with multiple phase singularity points, can be used for sensing applications. As a proof of concept, we showed that our lithography-free multilayered stack could be utilized as an ultrasensitive sensor susceptible to small refractive index changes from $CO_2$ gas flows.

Comparing our design with previously reported results on the generalized Brewster effect[26–29] and phase-change thin films based perfect absorber designs,[14,16,17] multiple obvious advantages can be seen for our approach. Our multilayered heterostructure stack exploits the strong interference effect, while most results rely on the Fabry-Perot (FP) resonant cavities for achieving strong/perfect absorption. It has been reported that the robust nature of strong interference systems with incidence angle and the reduced number of interfaces makes them attractive.[7,10] Moreover, our lithography-free layered thin film stacks have nanofabrication advantages compared with plasmonic surfaces requiring lithography steps. But more importantly, the novelty of our work relies on our ability to produce an optical device capable of achieving strong/perfect absorption for both s- and p- polarizations at the same Brewster angle. While the results reported in literature provide theoretical and experimental basis for potentially achieving the generalized Brewster effect, the proposed designs up to now did not support the coexistence of such vanishing reflectance values for both polarizations. In contrast, for our design the coexistence at a single Brewster angle of vanishing reflectances for both polarizations has been proven here with the additional degree of freedom that this coexistence is largely preserved when switching between the amorphous and crystalline phases of $Sb_2Se_3$.

## Conclusions

Our work demonstrates a multilayered heterostructure design capable of achieving strong/perfect absorption for both s- and p- polarized lights at a single light incidence angle. By designing our heterostructure stack carefully, we realized a thickness range where reflectance vanishes for both s- and p- polarized lights. In addition, we produced an optical device that experimentally validated the coexistence of points of darkness for both s- and p- polarized lights at an incidence angle of 70°. Vanishing reflectance values are associated with singular phase behaviors with abrupt changes at the points of darkness. We demonstrated the existence of two phase singularities in the as-deposited amorphous phase of the active phase-change material $Sb_2Se_3$, while additional two phase singularities are realized after switching to the crystalline phase. As a proof of concept, we showed that phase singularities are ultrasensitive to small deviations in optical constants, as we introduced here by a varying $CO_2$ flow, and could thus be utilized in various (sensor) applications. The tunability from phase-switching and the coexisting vanishing reflectance values for both s- and p- polarizations provide promising application options with versatile designs that are easily producible.

## Acknowledgments

This project has received funding from the European Union's Horizon 2020 Research and Innovation Programme "BeforeHand" (Boosting Performance of Phase Change Devices by Hetero- and Nanostructure Material Design)" under Grant Agreement No. 824957.

## References

(1) ElKabbash, M.; Iram, S.; Letsou, T.; Hinczewski, M.; Strangi, G.; ElKabbash, M.; Iram, S.; Letsou, T.; Hinczewski, M.; Strangi, G. Designer Perfect Light Absorption Using Ultrathin Lossless Dielectrics on Absorptive Substrates. *Adv. Opt. Mater.* **2018**, *6* (22), 1800672. https://doi.org/10.1002/ADOM.201800672.
(2) K. V., S.; ElKabbash, M.; Caligiuri, V.; Singh, R.; De Luca, A.; Strangi, G. Perfect Light Absorption in Thin and Ultra-Thin Films and Its Applications. **2019**, 3–27. https://doi.org/10.1007/978-981-13-8891-0_1.
(3) Yao, Y.; Liao, Z.; Liu, Z.; Liu, X.; Zhou, J.; Liu, G.; Yi, Z.; Wang, J. Recent Progresses

on Metamaterials for Optical Absorption and Sensing: A Review. *Journal of Physics D: Applied Physics*. 2021, p 21. https://doi.org/10.1088/1361-6463/abccf0.
(4) Cheng, F.; Gao, J.; Luk, S. T.; Yang, X. Structural Color Printing Based on Plasmonic Metasurfaces of Perfect Light Absorption. *Sci. Reports 2015 51* **2015**, *5* (1), 1–10. https://doi.org/10.1038/srep11045.
(5) Aydin, K.; Ferry, V. E.; Briggs, R. M.; Atwater, H. A. Broadband Polarization-Independent Resonant Light Absorption Using Ultrathin Plasmonic Super Absorbers. *Nat. Commun. 2011 21* **2011**, *2* (1), 1–7. https://doi.org/10.1038/ncomms1528.
(6) Li, Z.; Butun, S.; Aydin, K. Large-Area, Lithography-Free Super Absorbers and Color Filters at Visible Frequencies Using Ultrathin Metallic Films. *ACS Photonics* **2015**, *2*, 183–188. https://doi.org/10.1021/ph500410u.
(7) Song, H.; Guo, L.; Liu, Z.; Liu, K.; Zeng, X.; Ji, D.; Zhang, N.; Hu, H.; Jiang, S.; Gan, Q.; Song, H.; Liu, K.; Zeng, X.; Ji, D.; Zhang, N.; Gan, Q.; Guo, L.; Liu, Z.; Jiang, S.; Hu, H. Nanocavity Enhancement for Ultra-Thin Film Optical Absorber. *Adv. Mater.* **2014**, *26* (17), 2737–2743. https://doi.org/10.1002/ADMA.201305793.
(8) Sreekanth, K. V.; Sreejith, S.; Han, S.; Mishra, A.; Chen, X.; Sun, H.; Lim, C. T.; Singh, R. Biosensing with the Singular Phase of an Ultrathin Metal-Dielectric Nanophotonic Cavity. *Nat. Commun. 2018 91* **2018**, *9* (1), 1–8. https://doi.org/10.1038/s41467-018-02860-6.
(9) Kats, M. A.; Capasso, F. Optical Absorbers Based on Strong Interference in Ultra-Thin Films. *Laser Photonics Rev.* **2016**, *10* (5), 735–749. https://doi.org/10.1002/lpor.201600098.
(10) Kats, M. A.; Blanchard, R.; Genevet, P.; Capasso, F. Nanometre Optical Coatings Based on Strong Interference Effects in Highly Absorbing Media. *Nat. Mater.* **2013**, *12* (1), 20–24. https://doi.org/10.1038/nmat3443.
(11) Wuttig, M.; Bhaskaran, H.; Taubner, T. Phase-Change Materials for Non-Volatile Photonic Applications. *Nat. Photonics* **2017**, *11* (8), 465–476. https://doi.org/10.1038/nphoton.2017.126.
(12) Fan, Z.; Deng, Q.; Ma, X.; Zhou, S. Phase Change Metasurfaces by Continuous or Quasi-Continuous Atoms for Active Optoelectronic Integration. *Mater. 2021, Vol. 14, Page 1272* **2021**, *14* (5), 1272. https://doi.org/10.3390/MA14051272.
(13) Abdollahramezani, S.; Hemmatyar, O.; Taghinejad, H.; Krasnok, A.; Kiarashinejad, Y.; Zandehshahvar, M.; Alu, A.; Adibi, A.; Alù, A.; Alù, A.; Adibi, A. Tunable Nanophotonics Enabled by Chalcogenide Phase-Change Materials. *Nanophotonics* **2020**. https://doi.org/10.1515/nanoph-2020-0039.
(14) Cao, T.; Wei, C. W.; Simpson, R. E.; Zhang, L.; Cryan, M. J. Broadband Polarization-Independent Perfect Absorber Using a Phase-Change Metamaterial at Visible Frequencies. *Sci. Reports 2014 41* **2014**, *4* (1), 1–8. https://doi.org/10.1038/srep03955.
(15) Cueff, S.; Taute, A.; Bourgade, A.; Lumeau, J.; Monfray, S.; Song, Q.; Genevet, P.; Devif, B.; Letartre, X.; Berguiga, L. Reconfigurable Flat Optics with Programmable Reflection Amplitude Using Lithography-Free Phase-Change Material Ultra-Thin Films. *Adv. Opt. Mater.* **2021**, *9* (2), 2001291. https://doi.org/10.1002/ADOM.202001291.
(16) Tittl, A.; Michel, A.-K. U.; Schäferling, M.; Yin, X.; Gholipour, B.; Cui, L.; Wuttig, M.; Taubner, T.; Neubrech, F.; Giessen Tittl, H. A.; Schäferling, M.; Yin, X.; Neubrech, F.; Giessen, H.; Michel, A. U.; Wuttig, M.; Taubner, T.; Gholipour, B.; Cui, L. A Switchable Mid-Infrared Plasmonic Perfect Absorber with Multispectral Thermal Imaging Capability. *Adv. Mater.* **2015**, *27* (31), 4597–4603. https://doi.org/10.1002/ADMA.201502023.
(17) Wen, X.; Xiong, Q. A Large Scale Perfect Absorber and Optical Switch Based on Phase

Change Material (Ge2Sb2Te5) Thin Film. https://doi.org/10.1007/s40843-016-0129-7.
(18) Grigorenko, A. N.; Patskovsky, S.; Kabashin, A. V. Phase and Amplitude Sensitivities in Surface Plasmon Resonance Bio and Chemical Sensing. *Opt. Express, Vol. 17, Issue 23, pp. 21191-21204* **2009**, *17* (23), 21191–21204. https://doi.org/10.1364/OE.17.021191.
(19) Grigorenko, A. N.; Nikitin, P. I.; Kabashin, A. V. Phase Jumps and Interferometric Surface Plasmon Resonance Imaging. *Appl. Phys. Lett.* **1999**, *75* (25), 3917–3919. https://doi.org/10.1063/1.125493.
(20) Kravets, V. G.; Schedin, F.; Jalil, R.; Britnell, L.; Gorbachev, R. V.; Ansell, D.; Thackray, B.; Novoselov, K. S.; Geim, A. K.; Kabashin, A. V.; Grigorenko, A. N. Singular Phase Nano-Optics in Plasmonic Metamaterials for Label-Free Single-Molecule Detection. *Nat. Mater. 2013 124* **2013**, *12* (4), 304–309. https://doi.org/10.1038/NMAT3537.
(21) Berkhout, A.; Koenderink, A. F. Perfect Absorption and Phase Singularities in Plasmon Antenna Array Etalons. *ACS Photonics* **2019**, *6* (11), 2917–2925. https://doi.org/10.1021/ACSPHOTONICS.9B01019/ASSET/IMAGES/MEDIUM/PH9B01019_M010.GIF.
(22) Zeng, S.; Hu, S.; Xia, J.; Anderson, T.; Dinh, X.-Q.; Meng, X.-M.; Coquet, P.; Yong, K.-T. Graphene-MoS$_2$ Hybrid Nanostructures Enhanced Surface Plasmon Resonance Biosensors. *Sensors Actuators B* **2015**, *207*, 801–810. https://doi.org/10.1016/j.snb.2014.10.124.
(23) Oh, S. H.; Altug, H.; Jin, X.; Low, T.; Koester, S. J.; Ivanov, A. P.; Edel, J. B.; Avouris, P.; Strano, M. S. Nanophotonic Biosensors Harnessing van Der Waals Materials. *Nat. Commun. 2021 121* **2021**, *12* (1), 1–18. https://doi.org/10.1038/s41467-021-23564-4.
(24) Brewster, D. On the Laws Which Regulate the Polarisation of Light by Reflexion from Transparent Bodies. *Philos. Trans. R. Soc. Lond.* **1815**, *105*, 125–159. https://doi.org/10.1098/rstl.1815.0010.
(25) Thomas, P. A.; Menghrajani, K. S.; Barnes, W. L. All-Optical Control of Phase Singularities Using Strong Light-Matter Coupling. *Nat. Commun. 2022 131* **2022**, *13* (1), 1–6. https://doi.org/10.1038/s41467-022-29399-x.
(26) Paniagua-Domínguez, R.; Yu, Y. F.; Miroshnichenko, A. E.; Krivitsky, L. A.; Fu, Y. H.; Valuckas, V.; Gonzaga, L.; Toh, Y. T.; Kay, A. Y. S.; Lukyanchuk, B.; Kuznetsov, A. I. Generalized Brewster Effect in Dielectric Metasurfaces. *Nat. Commun. 2016 71* **2016**, *7* (1), 1–9. https://doi.org/10.1038/ncomms10362.
(27) Lavigne, G.; Caloz, C. Generalized Brewster Effect Using Bianisotropic Metasurfaces. *Opt. Express* **2021**, *29* (7), 11361. https://doi.org/10.1364/OE.423078.
(28) Ding, L.; Qiu, T.; Zhang, J.; Wen, X. Generalized Brewster Effect Tuned Optically in a Graphene/Substrate System. **2019**. https://doi.org/10.1088/2040-8986/ab4fa1.
(29) Sreekanth, K. V.; Elkabbash, M.; Medwal, R.; Zhang, J.; Letsou, T.; Strangi, G.; Hinczewski, M.; Rawat, R. S.; Guo, C.; Singh, R. Generalized Brewster Angle Effect in Thin-Film Optical Absorbers and Its Application for Graphene Hydrogen Sensing. *ACS Photonics* **2019**, *6* (7), 1610–1617. https://doi.org/10.1021/acsphotonics.9b00564.
(30) Delaney, M.; Zeimpekis, I.; Lawson, D.; Hewak, D. W.; Muskens, O. L. A New Family of Ultralow Loss Reversible Phase-Change Materials for Photonic Integrated Circuits: $Sb_2S_3$ and $Sb_2Se_3$. *Adv. Funct. Mater.* **2020**, *30* (36), 2002447. https://doi.org/10.1002/adfm.202002447.
(31) Faneca, J.; Zeimpekis, I.; Ilie, S. T.; Bucio, T. D.; Grabska, K.; Hewak, D. W.; Gardes, F. Y. Towards Low Loss Non-Volatile Phase Change Materials in Mid Index Waveguides. *Neuromorphic Comput. Eng.* **2021**, *1* (1), 014004. https://doi.org/10.1088/2634-4386/ac156e.

(32) Dong, W.; Liu, H.; Behera, J. K.; Lu, L.; H Ng, R. J.; Valiyaveedu Sreekanth, K.; Zhou, X.; W Yang, J. K.; Simpson, R. E.; Dong, W.; Liu, H.; Behera, J. K.; Lu, L.; H Ng, R. J.; Zhou, X.; W Yang, J. K.; Simpson, R. E.; Sreekanth, K. V. Wide Bandgap Phase Change Material Tuned Visible Photonics. *Adv. Funct. Mater.* **2019**, *29* (6), 1806181. https://doi.org/10.1002/ADFM.201806181.

(33) Yimam, D. T.; Liang, M.; Ye, J.; Kooi, B. J. 3D Nanostructuring of Phase-Change Materials Using Focused Ion Beam Towards Versatile Optoelectronics Applications. Advanced Materials 2023, 2303502. https://doi.org/10.1002/ADMA.202303502.

(34) Yimam, D. T.; Kooi, B. J. Thickness-Dependent Crystallization of Ultrathin Antimony Thin Films for Monatomic Multilevel Reflectance and Phase Change Memory Designs. *ACS Appl. Mater. Interfaces* **2022**, *14* (11), 13593–13600. https://doi.org/10.1021/acsami.1c23974.

(35) Yimam, D. T.; Ahmadi, M.; Kooi, B. J. Van Der Waals Epitaxy of Pulsed Laser Deposited Antimony Thin Films on Lattice-Matched and Amorphous Substrates. Mater Today Nano 2023, 23, 100365. https://doi.org/10.1016/J.MTNANO.2023.100365..

# Supplementary Information for

# Coupling phase-switching with generalized Brewster effect for tunable optical sensor designs

Daniel T. Yimam*, Dennis van der Veen, Teodor Zaharia, Maria Loi, Bart J. Kooi*

Zernike Institute for Advanced Materials, University of Groningen, Nijenborgh 4, 9747 A Groningen, The Netherlands

*Corresponding authors. Email: d.t.yimam@rug.nl, b.j.kooi@rug.nl

SI 1 – Reflectance for crystalline $Sb_2Se_3$ films

The crystallization of the active $Sb_2Se_3$ layer produces reflectance profiles different from what we show in Fig. 2 of the main text. The optical constants difference for the as-deposited amorphous and crystalline phases of the $Sb_2Se_3$ layer, as seen in Fig. 1b of the main text, is responsible for the difference in reflectance profiles. Nevertheless, reflectance vanishing points for both p- and s- polarized lights also occur for the crystalline $Sb_2Se_3$ films as can be seen in Fig. S1a and S1b. Compared with the as-deposited amorphous phase in Fig. 2, the points of darkness for the crystalline phase are shifted towards higher wavelength values. From the profiles in Fig. S1c and S1c, we can see significantly large ratio values for a layer thickness of 23.5 nm. However, the crystallization of phase-change layers is associated with a reduced thickness (from increased density), and thus we also have to consider a lower thickness value (≈22 nm). Nonetheless, coexisting and large resonance peaks are expected for both thickness values.

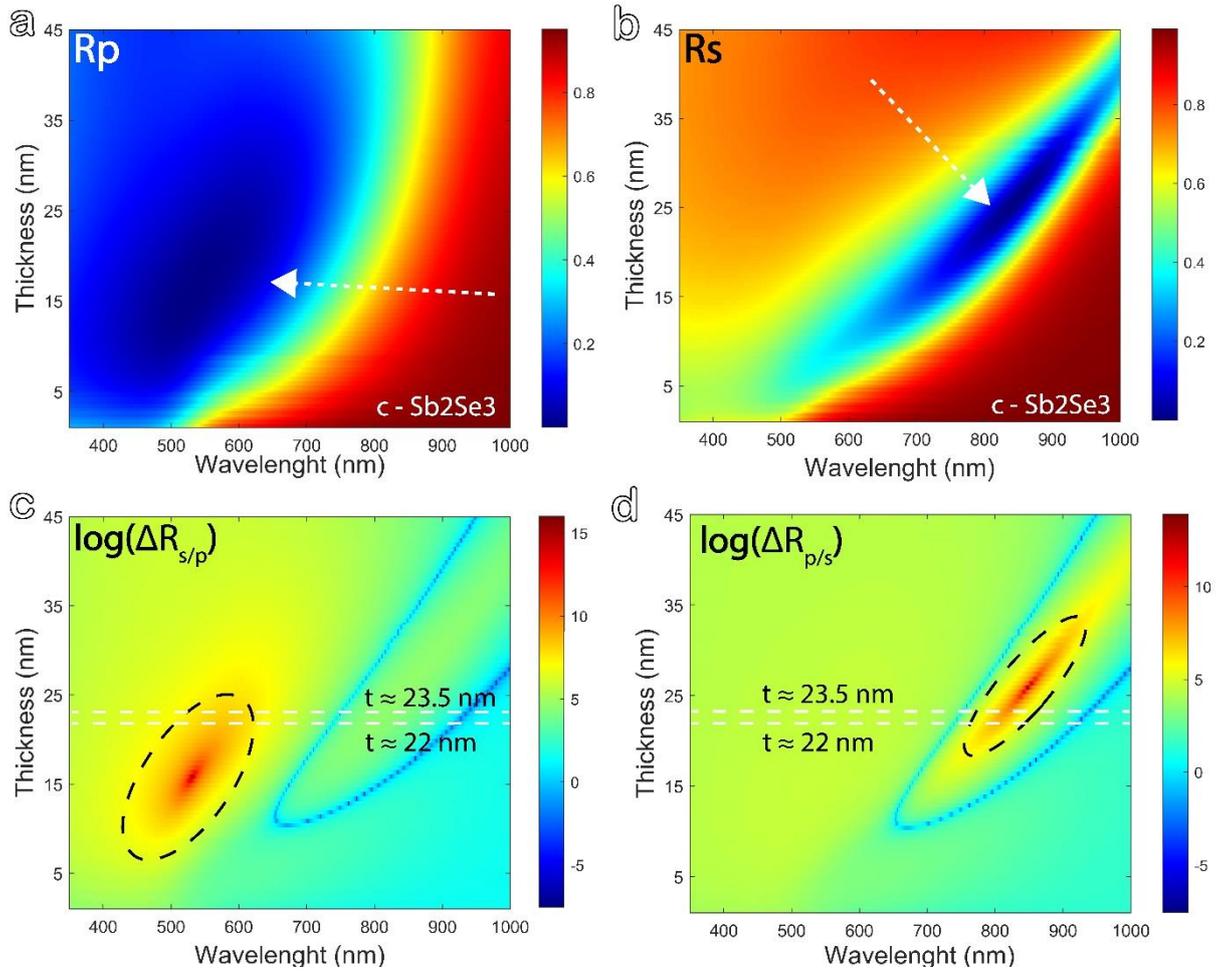

**Figure S1**. Calculated s- and p- polarization reflectance vs. thickness for the crystalline phase of the active $Sb_2Se_3$ layer in the heterostructure device. In (a) and (b), the polarization-dependent reflectance profiles are presented for varying thicknesses of the crystalline $Sb_2Se_3$ layer. The layer thickness was varied from 0 – 45 nm, and the calculations were done for the wavelength range of 350 – 1000 nm. The white arrows indicate minimum reflectance values for both s- and p- polarized lights. In (c) and (d), the log values of the polarization ratios (Rs/Rp and Rp/Rs) are given. The largest ratio values can be identified for a wide range of layer thickness values.

## SI 2 – Crystalline $Sb_2Se_3$ (23.5 nm)

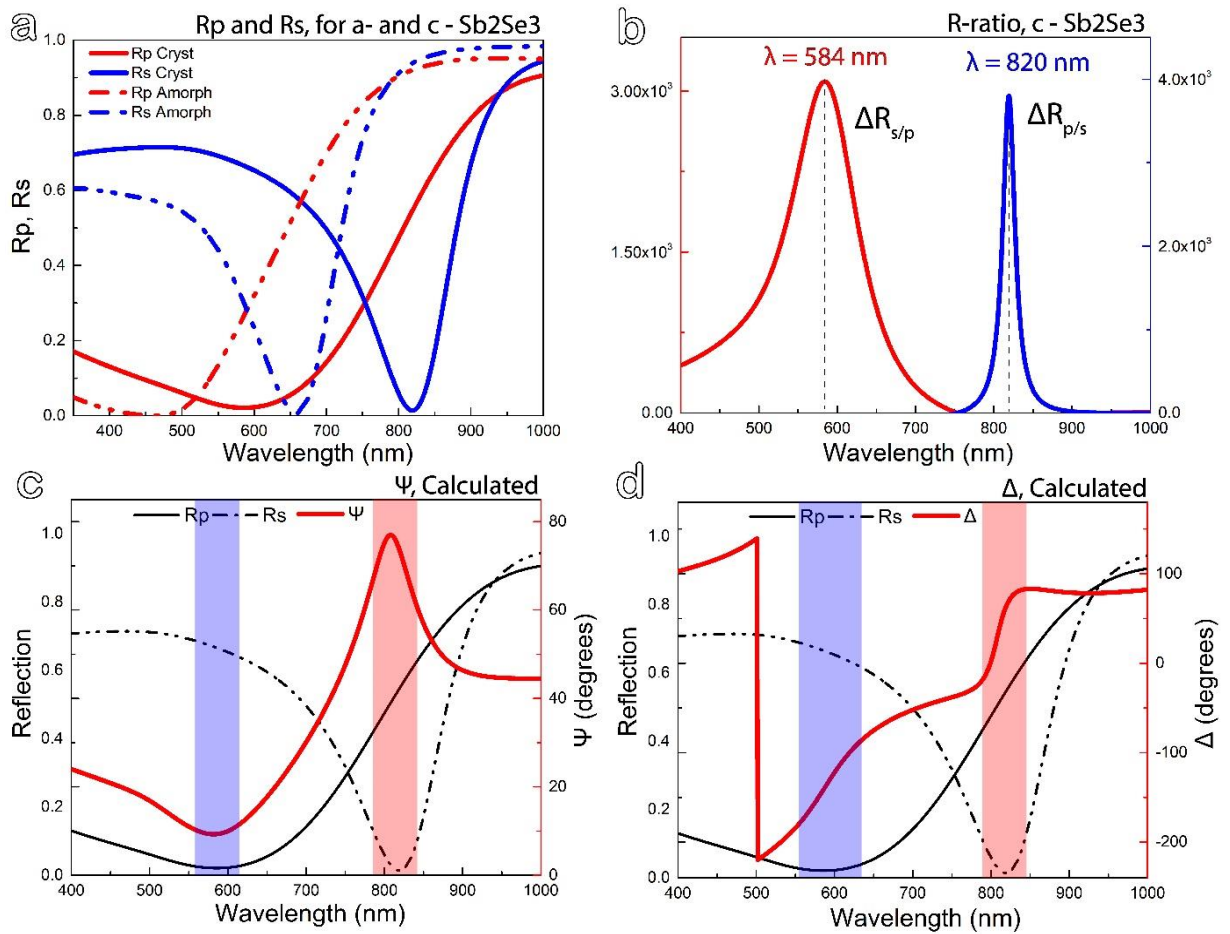

**Figure S2**. The polarization-dependent reflectance profiles, the relative reflectance ratios, and the associated phase singularities. (a) The calculated s- and p- polarized light reflectance profiles for a 23.5 nm crystalline $Sb_2Se_3$ layer at a 70° incidence angle. (b) The relative reflectance ratios between polarizations. The calculated ellipsometry parameters (c) ψ and (d) Δ, with the s- and p- polarization reflectance values overlayed.

The reflectance values for both the as-deposited amorphous and crystalline phases of $Sb_2Se_3$ are presented in Fig. S2a. Here the layer thickness is fixed to 23.5 nm. Two crucial informations are visible in the figure. The first information is the deviation in reflectance values for the as-deposited amorphous and crystalline phases for both polarizations. The reflectance values are red-shifted towards higher wavelength values for the crystalline phase. The second and most important information is the strong/perfect absorption present for both phases. For a 23.5 nm thick crystalline $Sb_2Se_3$ layer, reflectance values of Rp = 2% and Rs = 1.4% were registered at the two points of darkness. Although such reflectance values are relatively high compared with the as-deposited amorphous phase of exact thickness, the values still translate to strong light absorption and, thus, phase singularities. The presence of reflectance vanishing points for both phases provides an opportunity to move towards tunable optical sensor designs. The resonance peaks in Fig. S2b (at 584 nm and 820 nm for Rs/Rp and Rp/Rs, respectively) indicate the presence of phase singularities for the crystalline $Sb_2Se_3$ layer. Such resonance peaks for the amorphous phase were seen at 470 nm and 652 nm. To confirm the existence of phase singularities at the points of darkness, we calculated the ellipsometry ψ and Δ parameters. The abrupt changes in the ellipsometry parameters are indeed seen in Fig. S2c and S2d for both ψ and Δ, exactly at the wavelength values where the resonances are present.

## SI 3 – Crystalline $Sb_2Se_3$ (22 nm)

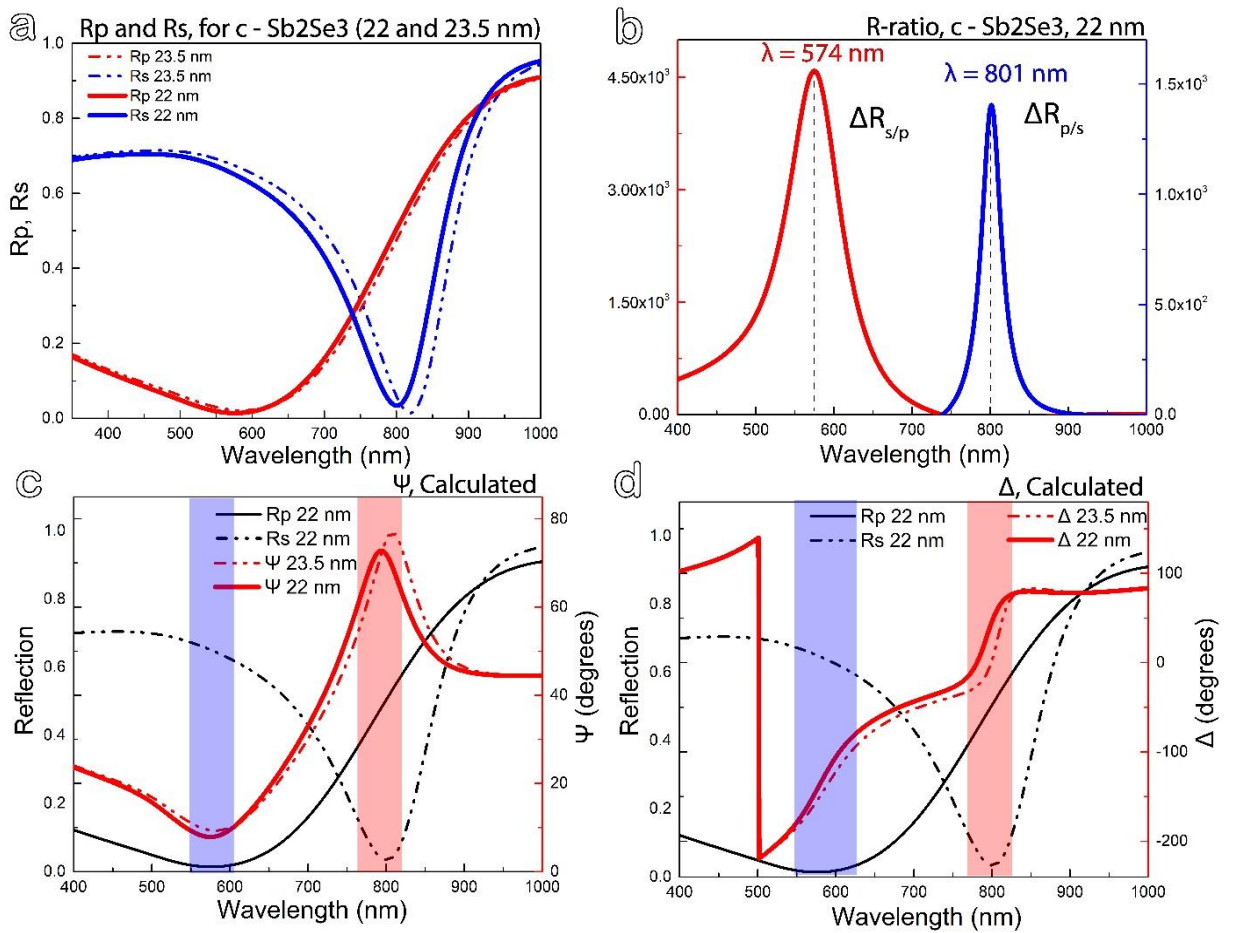

**Figure S3**. Reflectance and ratio value shifts for thickness change due to crystallization. (a) Calculated reflectance values for layer thickness values of 22 nm and 23.5 nm. The relatively small thickness value change caused a shift in the points of darkness towards lower wavelength values. The shift in wavelength value and the relative ratio changes are presented in (b). The resonance peaks shifted in wavelength values from 584 nm to 574 nm for Rs/Rp and from 820 nm to 801 nm for Rp/Rs. Moreover, the relative ratio values for Rs/Rp increased wheras they decreased for Rp/Rs.

As mentioned above, the crystallization of phase-change materials is accompanied by an increase in density. This density change (usually between 6 - 10 % for GST225 and $Sb_2Te_3$) manifests as a small layer thickness decrease in the crystalline phase compared to the amorphous phase. The thickness reduction has to be incorporated into the heterostructure device design. It is important to evaluate the relatively small thickness deviations since the strong interference effect is highly sensitive, and such a small thickness change can still cause deviations in the reflectance profiles. We recalculated the reflectance values for both polarizations considering a crystalline $Sb_2Se_3$ layer thickness of 22 nm to evaluate the effect of such thickness deviation from phase switching. We compared it with the results presented in SI2 for a thickness value of 23.5 nm.

The reflectance profiles for $Sb_2Se_3$ layer thickness values of 22 nm and 23.5 nm are presented in Fig. S3a. Both polarizations show small shifts in reflectance values. To extract the exact shifted values, we plot the reflectance ratios for the 22 nm layer thickness in Fig. S3b. The effect of the small thickness deviation can be seen from the value changes in both resonance peaks and the associated wavelength values. The layer thickness change (from 23.5 nm to 22 nm) is associated with a wavelength value shift for the resonance peaks (from 584 nm to 574 nm for Rp and from 820 nm to 801 nm for Rs).

Moreover, the thickness change also changed the ratio values where Rs/Rp increased while Rp/Rs reduced. We calculated the ellipsometry $\psi$ and $\Delta$ parameters for a thickness value of 22 nm to see the effect of such thickness change on the phase singularity. The $\psi$ and $\Delta$ parameters for a thickness of 22 nm and 23.5 nm for comparison are plotted in Fig. S3c and S3c. Comparable phase singularities can be seen for both thickness values, with only a small shift in wavelength values.

## SI 4 – $\psi$ vs. gas flow rate

Fig. 5 of the main text presents the ellipsometry $\Delta$ parameter for $CO_2$ gas flow rate changes. The measurements were done for a wavelength range of 480 – 490 nm, corresponding to one of the resonance peaks (thus phase singularity). In the corresponding wavelength ranges, the ellipsometry $\psi$ parameter was also collected. Fig. S4 presents the correlation between the measured $\psi$ parameter and the gas flow rate. In Fig. S4a and S4b, the $\psi$ values for all gas flow rates and a zoomed-in image are depicted, respectively. Although relatively small, the $\psi$ values shift towards smaller values for increasing gas flow rate. The relative $\psi$ value changes are plotted in Fig. S4c. Again, somewhat linear and downward trends can be observed. Similar to the $\Delta$ values, the $\psi$ parameter also shows a linearly dependent correlation with gas flow rate, as seen in Fig. S4d for a wavelength value of 486 nm.

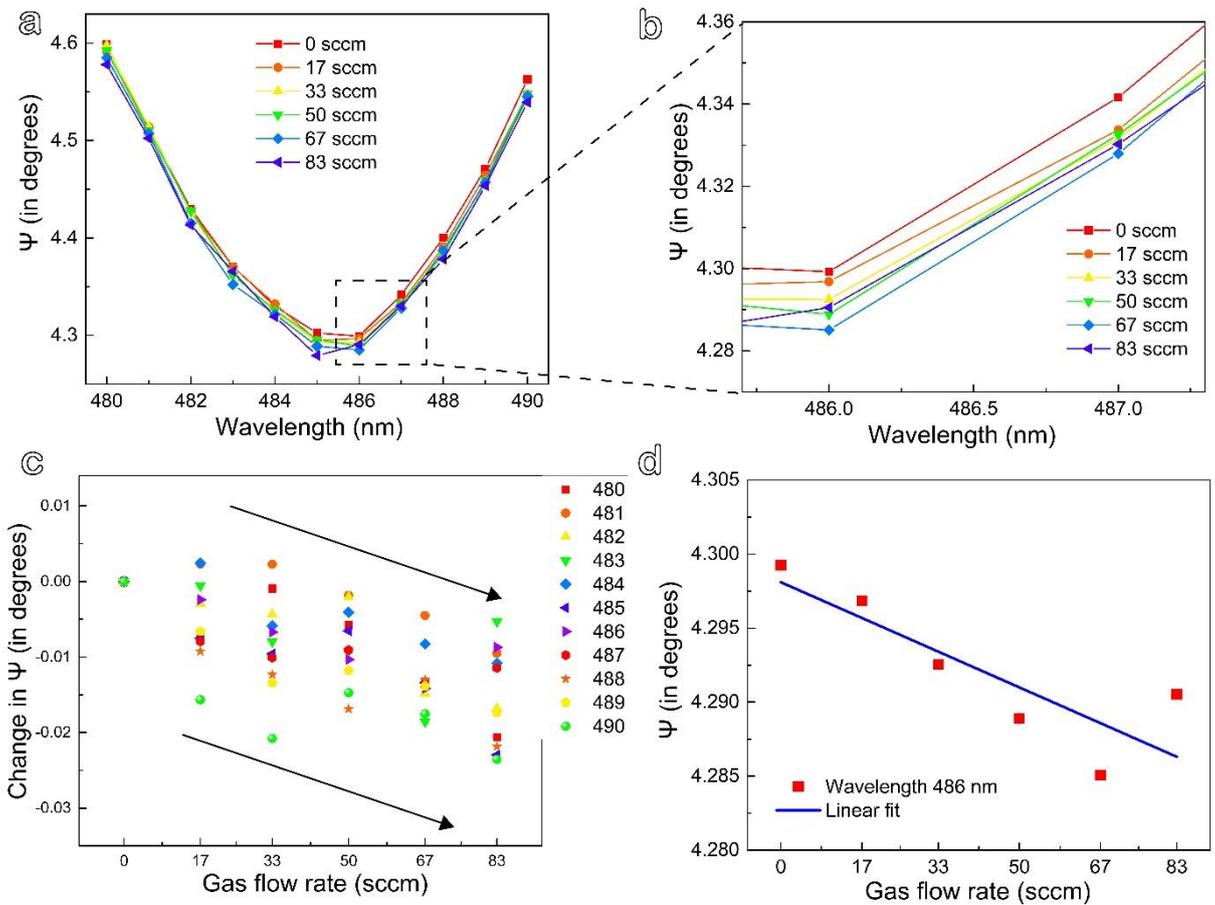

**Figure S4**. Gas sensing experiment results for the wavelength range of 480 – 490 nm. (a) The ellipsometry ψ parameter collected for gas flow rates changing from 0 – 83 sccm. (b) A zoomed-in image of some of the ψ values showing systematic changes with respect to the gas flow rate. (c) The relative change in ψ values, with and without gas flow for each rate, shows a linear dependence on gas flow rate values. (d) The ψ value change with the gas flow rate for a single wavelength value of 486 nm, showing a linear correlation.